\documentclass[preprint]{aastex}
\newcommand{\com}[1]{}
\usepackage{amsmath,amssymb,amsfonts}
\usepackage{graphicx}
\usepackage[update,prepend]{epstopdf}
\usepackage{hyperref}
\usepackage{array}
\usepackage{cite}
\usepackage{natbib}
\usepackage{morefloats}
\usepackage{color,soul}

\title{Simultaneous Observations of Giant Pulses from Pulsar PSR~B0950+08 at 42~MHz and 74~MHz}

\author{
Jr-Wei Tsai\altaffilmark{1},
John H. Simonetti\altaffilmark{1},
Bernadine Akukwe\altaffilmark{1}, 
Brandon Bear\altaffilmark{1},
Jonathan D. Gough\altaffilmark{2},
Peter Shawhan\altaffilmark{3},
Michael Kavic\altaffilmark{4}
}

\altaffiltext{1}{Department of Physics, Virginia Tech, Blacksburg, VA 24061, U.S.A}
\altaffiltext{2}{Department of Chemistry, Lehman College,
Bronx, NY 10468}
\altaffiltext{3}{Department of Physics, University of Maryland, College Park, MD 20742, USA}
\altaffiltext{4}{Department of Physics, Long Island University, Brooklyn, New York 11201, U.S.A.}

\begin{abstract}
We report the detection of giant pulse emission from PSR~B0950+08 in 12 hours of observations made simultaneously at 42~MHz and 74~MHz, using the first station of the Long Wavelength Array, LWA1.  We detected 275 giant pulses (in 0.16\% of the pulse periods) and 465 giant pulses (0.27\%) at 42 and 74~MHz, respectively. The pulsar is weaker and produces less frequent giant pulses than at 100~MHz. Here, giant pulses are taken as having $\geq$ 10 times the flux density of an average pulse; their cumulative distribution of pulse strength follows a power law, with a index of $-$4.1 at 42~MHz and $-$5.1 at 74~MHz, which is  much less steep than would be expected if we were observing the tail of a Gaussian distribution of normal pulses.  We detected no other transient pulses in a wide dispersion measure range from 1 to 5000~pc~cm$^{-3}$.  There were 128 giant pulses detected within in the same periods from both 42 and 74~MHz, which means more than half of them are not generated in a wide band.  We use CLEAN-based algorithm to analyze the temporal broadening and conclude that the scattering effect from the interstellar medium can not be observed.  We calculated the altitude $r$ of the emission region using the dipolar magnetic field model. We found $r$(42~MHz) = 29.27~km ($0.242\%$ of $R_{LC}$) and $r$(74~MHz) = 29.01~km ($0.240\%$ of $R_{LC}$) for the average pulse, while for giant pulses, $r$(42~MHz) = 29.10~km ($0.241\%$ of $R_{LC}$) and $r$(74~MHz) = 28.95~km ($0.240\%$  of $R_{LC}$).  Giant pulses, which have a double-peak structure, have a smaller mean peak-to-peak separation compared to the average pulse.

\end{abstract}


\newpage

\begin{document}

\section{Introduction}
One year after the discovery of the first pulsar in 1967 \citep{1968Natur.217..709H}, the first giant pulse (GP) was detected from the Crab pulsar (PSR B0531+21) by \citet{1968Sci...162.1481S}. GPs have an observed flux density that is tens or hundreds of times larger than for an average pulse (AP). Observations of GPs can illuminate the underlying pulsar pulse production mechanism and can serve as an effective probe of the interstellar medium (ISM). Moreover, simultaneous observations at multiple frequencies allow for a more nuanced investigation of these phenomena. In the current work we perform such an analysis using simultaneous observation of PSR~B0950+08 at 42 and 74 MHz. These observations were conducted using the first station of the Long Wavelength Array (LWA1) \citep{2013ITAP...61.2540E}. 

Only a small minority of pulsars emit GPs.  Initially, all pulsars observed to emit GPs were found to possess a very strong magnetic field in the region of the pulsar's light cone, with $B_{LC}>10^5$ G \citep{1996ApJ...457L..81C}. This changed with the observations of PSR~B0950+08 by \citet{2001Ap&SS.278...61S}, indicating that pulsars with a lower $B_{LC}$ (150 G for PSR~B0950+08) can also emit giant pulses.  Another common property of GPs is that they exhibit a non-Gaussian power-law distribution in the peak flux density.  GP observations of PSR~B0950+08 over the frequency range 39-111~MHz  indicate further that the rate and strength of GP emission is reduced at $\sim$39~MHz as compared to $\sim$100~MHz \citep{2012AJ....144..155S, 2012ARep...56..430S, 2015AJ....149...65T}. In this paper we extend the previous study of PSR~B0950+08 by \citet{2015AJ....149...65T} with simultaneous observations at two frequencies of 42 and 74~MHz. We detected 275 and 465 GPs at 42 and 74~MHz, respectively. The rate and strength of these GPs are less than has been observed at $\sim$100~MHz. This indicates that this pulsar is weaker and produces less frequent GPs below 100~MHz, than above 100~MHz. We found the AP has a width which is consistent with previous observations. We detected no other transient pulses with signal-to-noise ratio (SNR) $>7$ over a wide range of dispersion measures range from 1 to 5000~pc~cm$^{-3}$.

Simultaneous observations of pulses at multiple frequencies can be used to investigate the pulsar's beam structure.  In particular it is possible to analyze the emission altitude difference for the simultaneous observations.  We can infer emission altitude from the peak-to-peak separation of double-peaked pulses at different frequencies, assuming a dipolar magnetic field of the pulsar. We determined the emission altitude in the context of a dipolar magnetic field for our set of observations for both GPs and APs.  
We found $r$(42~MHz) = 29.27~km ($0.242\%$ of $R_{LC}$) and $r$(74~MHz) = 29.01~km ($0.240\%$ of $R_{LC}$) for the average pulse, while for giant pulses, $r$(42~MHz) = 29.10~km ($0.241\%$ of $R_{LC}$) and $r$(74~MHz) = 28.95~km ($0.240\%$  of $R_{LC}$). The $R_{LC}$ is the light cylinder radius of this pulsar. The mean peak-to-peak separation of GPs is smaller than APs for both observing frequencies.

Observations of pulsars have been used to probe the spatial spectrum of irregularities in the ISM. Temporal pulse widths, at different frequencies, can be used to determine the spectral index of a power law describing the interstellar irregularities. The most commonly discussed power law is for a Kolmogorov spectrum. Our previous observations of PSR~B0950+08 suggested that the spectrum along this line-of-sight deviates from Kolmogorov \citep{2015AJ....149...65T}. In this work, we explore this issue further by considering the effect of scattering on the GPs and APs simultaneous observed at two frequencies. To analyze the scattering effect we made use of the CLEAN-based algorithm \citep{2003ApJ...584..782B} to determine the exponential decay time for APs and pulses with SNR $>7$, assuming a thin screen scattering model.  We found different exponential decay time from pulses with SNR$>7$ (2.5$\pm1.3$/1.1$\pm0.7$~ms for 42/74~MHz) and AP (5.4/3.4~ms for 42/74~MHz).  Because the scattering effect from the ISM should be the same for individual pulses or average profiles, there should be other effects that dominate the pulse broadening, so we still can not determine the scattering time of PSR~B0950+08 for frequency down to 42~MHz.

We detail our observations and data reduction in section \ref{obanddr}.  Next we explain the scattering model used as well as the method for determining flux densities.  We describe the pulse behavior from observation in section \ref{behaviorofpsr}.  We then discuss the interpretation of these results in the context the cone structure beam model in section \ref{sect:altitudes} and scattering in section \ref{sect:scattering}. Finally we summarize our main findings in section \ref{sect:conclusion}.

\section{Observations and Data Reduction}\label{obanddr}
LWA1 \citep{2011ITAP...59.1855E} is a radio telescope array operating in the frequency range 10--88~MHz, located in central New Mexico.  The telescope consists of 256 dual-polarized dipole antennas distributed over an area of about 110~m by 100~m, plus 5 outliers at distances of 200--500 m from the core of the array, for a total of 261 dual-polarized antennas.  The outputs of the dipoles are individually digitized and can be formed into beams (DRX beam-forming mode).  Four fully independent dual-polarization beams capable of pointing anywhere in the sky are available; each beam has two independent frequency tunings (selectable from the range 10--88~MHz) with bandwidths of up to about 17~MHz.  The full-width at half-maximum (FWHM) beamwidth for zenith pointing is approximately $4.3^\circ$ at 74~MHz and depends on observing frequency as $\nu^{-1.5}$.  The system temperature is dominated by the Galactic emission and so the beam sensitivity of the instrument is dependent on the LST of the observation and the direction of the beam. The ability of the LWA1 to observe two frequencies simultaneously provides a powerful tool for studying radio sources.  The simultaneity of observations at different frequencies is useful for studying the profile frequency dependence and the temporal broadening of pulses \citet{2013ITAP...61.2540E}.  

Observations of PSR~B0950+08 were conducted using LWA1 in beam forming mode from January 25 through January 27, 2014. The observation set is composed of 4 consecutive hours of data from each day beginning 2 hours before the pulsar passed the meridian.  Observations were made at two frequencies simultaneously, centered at 42 and 74~MHz, each with a bandwidth of 16~MHz (due to low sensitivity). Two polarizations were recorded.

Using routines from the LWA Software Library (LSL) \citep{Dowell:2012rt}, we performed a 4096-channel Fast Fourier Transform (FFT) on each 0.209 ms of raw data, dividing the 19.6 MHz observing bandwidth into channels of 4.785 kHz.  Radio frequency interference (RFI) mitigation was performed on the data set, using the following procedure.  First, we obtained the average spectrum for each 2.09~second interval (each set of 10,000 consecutive spectra).  Next, we fit a 16th order polynomial to the 2.09~second average spectrum, and divided that average spectrum by the polynomial. The 16th order polynomial is the lowest order that fit the approximately 16 ripples in the bandpass, without undue suppression of narrow-band RFI. Finally, any frequency bin in the 2.09~second average spectrum that was greater than 3$\sigma$ above the mean was masked as RFI in all the corresponding 0.209~ms spectra.

The shape of the observed bandpass is not constant in time, and this variation must be removed to allow for an effective search for transient pulses.  The variation in the spectrum is dominated by the diurnal variation of the Galactic background.  Once the RFI-contaminated frequency bins were identified and masked, we determined and removed the varying shape of the bandpass using the following procedure.  First, for each 2.09 s of data, we computed a median spectrum for the 10,000 RFI-masked spectra.  Then, we used 150 such spectra to compute the median spectrum for approximately 5 minutes of observations. The 5 minute duration was chosen so the diurnal variation of the Galactic background was effectively smoothed out.  To further smooth the 5~minute spectrum, we performed a moving boxcar average across the 5~minute median-spectrum; the boxcar length used was 101 frequency channels.  Finally, we divided each 0.209~ms spectrum by the boxcar-smoothed spectrum corresponding to its epoch.  To prepare for further analysis we removed the first 360 channels and the last 395 channels from each spectrum, removing any end effects, and leaving a final bandwidth of 16~MHz.  The final spectra were arranged into spectrograms of frequency (vertical axis) and time (horizontal axis).

\citet{2003ApJ...596.1142C} describe in detail a technique suitable for searching for individual pulses of various origins, including pulsar GPs, in time-frequency data such as ours.  As a first look, we used this method on our data, taken in chunks of 5~minute duration. In essence, the technique consists of constructing dedispersed time series for a range of candidate DMs and smoothing each individual time series with effectively larger and larger averaging-boxcars to search for pulses of temporal width matched to the smoothing time --- which yields the best SNR for a candidate pulse.  Pulses of strengths and numbers larger than expected for the (assumed) Gaussian noise in our data are pulses of possible astrophysical origin.  (Another possibility is that they are RFI or other transient non-astrophysical events, but candidate pulses of the correct DM for PSR~B0950+08 are more likely to be our sought-after pulsar pulses.)  We performed incoherent dedispersion (summing intensities) in our spectrograms. We searched through time series for 28,451 candidate DMs in this manner, ranging from 1 to 5000 pc cm$^{-3}$ with a DM spacing $\delta$DM $= 0.0003$ DM.  The time series was smoothed in steps by averaging a moving boxcar of width equal to 2 time samples, and then removing one of the resulting time samples.  Repeated smoothing and decimating in this manner efficiently produces a set of time series of increasing smoothness. At each smoothing and decimation step the resulting time series are searched for pulses.  We performed 15 such steps for each dedispersed time series. Thus, the final time sample duration in the last-smoothed time series was $2^{15}\times0.2089~{\rm ms} = 6.85$~s.

In the entire search we found all transient events in the resulting time series with SNRs $\gtrsim$ 6.5 were for a DM of 2.97~pc~cm$^{-3}$, the DM for PSR~B0950+08.  No such strong pulses were found at other DMs.  Furthermore, the expected number of transient events due to Gaussian noise matches well our numbers of events at SNR$<$6.5, but events of SNR$>$6.5 are more numerous than expected from Gaussian noise alone, and appear at the DM of the pulsar.  Thus we are confident that by focusing on transient events that have a SNR$>$6.5, determined through this procedure, we are selecting pulses produced by PSR~B0950+08.  

We note here that the SNR determined for a pulse by the Cordes-McLaughlin procedure is computed in the time series smoothed to the temporal width of the pulse. This is a precise means of quantifying the SNR of a temporally-isolated \textit{single, dispersed pulse}, and, as such, is perhaps reasonable for describing anomalously intense pulses or GPs, which tend to be isolated.  But, it should be noted that quoted pulsar flux densities are averages in time, including both energy received during pulses, \textit{and} between pulses, i.e., effectively zero.  Thus, when we measure the flux density of our GPs and compare them to APs, we will adopt the more conventional time-average throughout a pulse period.

\section{Behavior of Pulses}\label{behaviorofpsr}
To study the flux density and phase of pulses we fit multiple Gaussians to both APs and GPs as both exhibited multiple-peak profiles.  The multiple Gaussian fits at both frequencies enable a simple comparison of the pulses at both frequencies.  Because the two observing frequencies are recorded simultaneously the relative phase can be determined after dedispersion.  We dedispersed the two frequencies independently using the same DM but produced separate dedispersed time series.

The flux density and phase of peaks were determined from the fitted Gaussians.  The flux density is calculated by converting the area under the fitted curves to the system equivalent flux density (SEFD). The flux density ratio of a GP and the AP is the ratio of the area under the fitted curves, while the area of the AP was divided by the number of pulses folded.

\subsection{Flux density}
Our observations did not include any drift scans on other objects for calibration, so we obtain rough flux densities for the AP and GPs using an estimated SEFD.  The SEFD is the flux density a source in the beam needs to produce a SNR of unity, for an observation of 1~Hz bandwidth and integration time of 1~second.  At low frequencies, the Galactic noise is the dominant contribution to system noise.  Ellingson established a rough model for estimating the SEFD, which takes account of the combined effects of all sources of noise \citep{Ellingsonsen}.  Ellingson uses a spatially uniform sky brightness temperature $T_B$ in his model, dependent on observing frequency $\nu$, where
\begin{equation}
T_b = 9751 \mbox{K} \left(\frac{\nu}{38\mbox{MHz}}\right)^{-2.55} 
\label{eqn:brightness temperature}
\end{equation}
and ignores the ground temperature contribution as negligible.  The receiver noise is about 250~K, but has little influence on the SEFD.  This model when applied to LWA1 shows that the correlation of the Galactic noise between antennas significantly desensitizes the array for beam pointings that are not close to the zenith.  It is also shown that considerable improvement is possible using beam-forming coefficients that are designed to optimize SNR under these conditions.  \citet{2013ITAP...61.2540E} checked this model with observations of strong flux density calibrators, finding the results roughly correct.  Based on the model of Ellingson and his drift scan results, and given our observations at transit are for non-zero zenith angle, we can estimate an appropriate SEFD to use for our observations, with an uncertainty of roughly 50\%.  The SNR of pulses away from the moment of transit are corrected by a factor which compensates for decreasing effective collecting area and increasing SEFD, with increasing zenith angle.

Thus, the flux density we assign to a pulse, as averaged across the entire pulse period, is
\begin{equation}
S = \frac{\rm SEFD}{\sqrt{2B\Delta t}}\ \frac{1}{N_{\rm bins}} \sum_{i=1}^{N_{\rm bins}} \frac{I_i}{\rm rms} = \frac{\rm SEFD}{\sqrt{2B\Delta t}}\ {\rm \overline{SNR}}
\label{eqn:flux density}
\end{equation}
where $B$ in~Hz is the bandwidth, $\Delta t$ in seconds is the duration of a time sample, $N_{\rm bins}$ is the number of time samples (bins) in a pulse period, the sum is over the full pulse period, the $I_i$ are the intensity values (arbitrary units) in the Gaussian pulse profile fitted to a pulse (a baseline average was already subtracted from the data), rms is measured in the baseline, and the $\rm \overline{SNR}$ is the average SNR during the pulse period.  Using \citet{2013ITAP...61.2540E} we assume the same SEFD of 15,000~Jy ($\pm$50\%) for both observing frequencies while transiting at a zenith angle of about 26$^\circ$.

\begin{figure}
\begin{center}
\includegraphics[width=.45\textwidth]{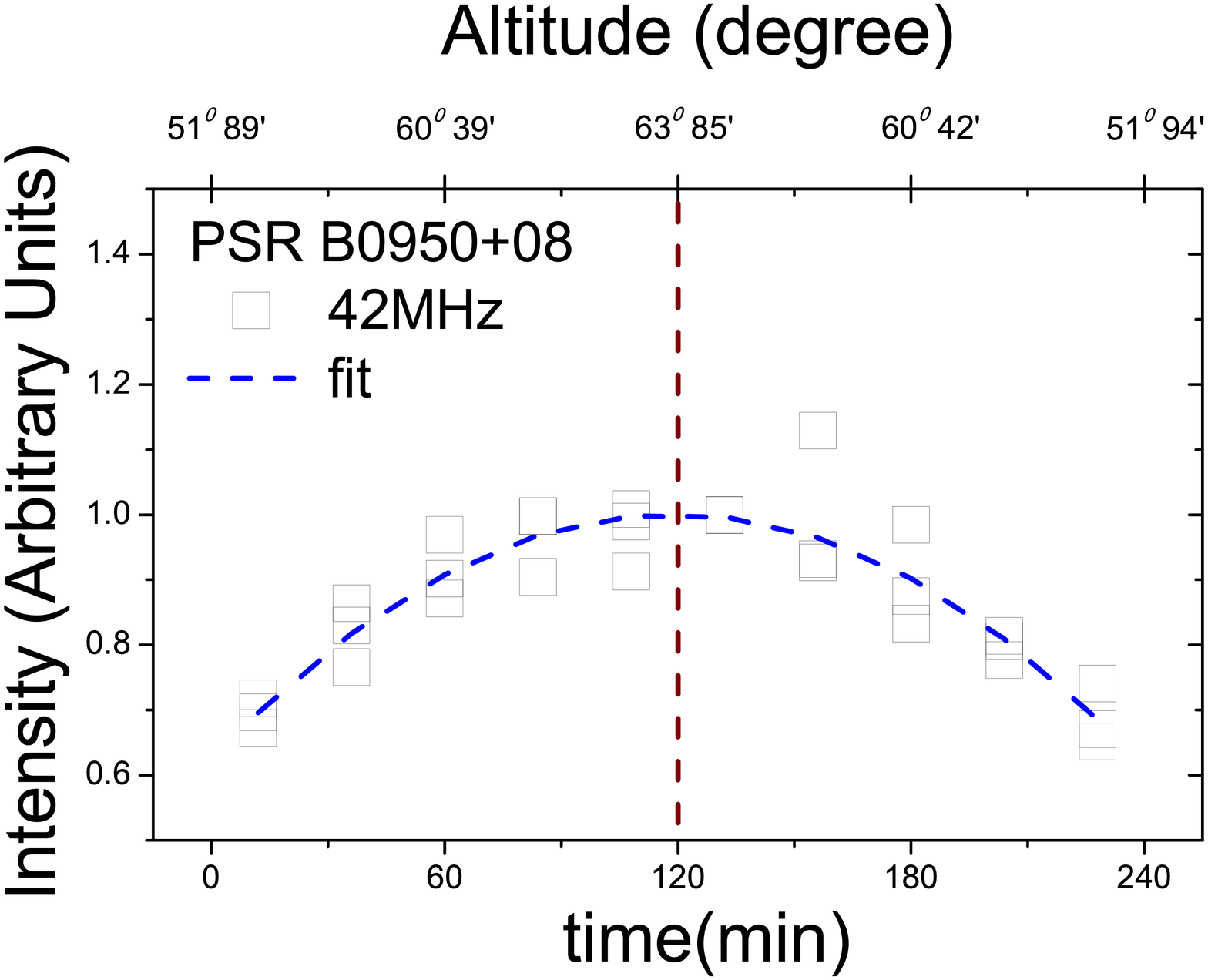}
\includegraphics[width=.45\textwidth]{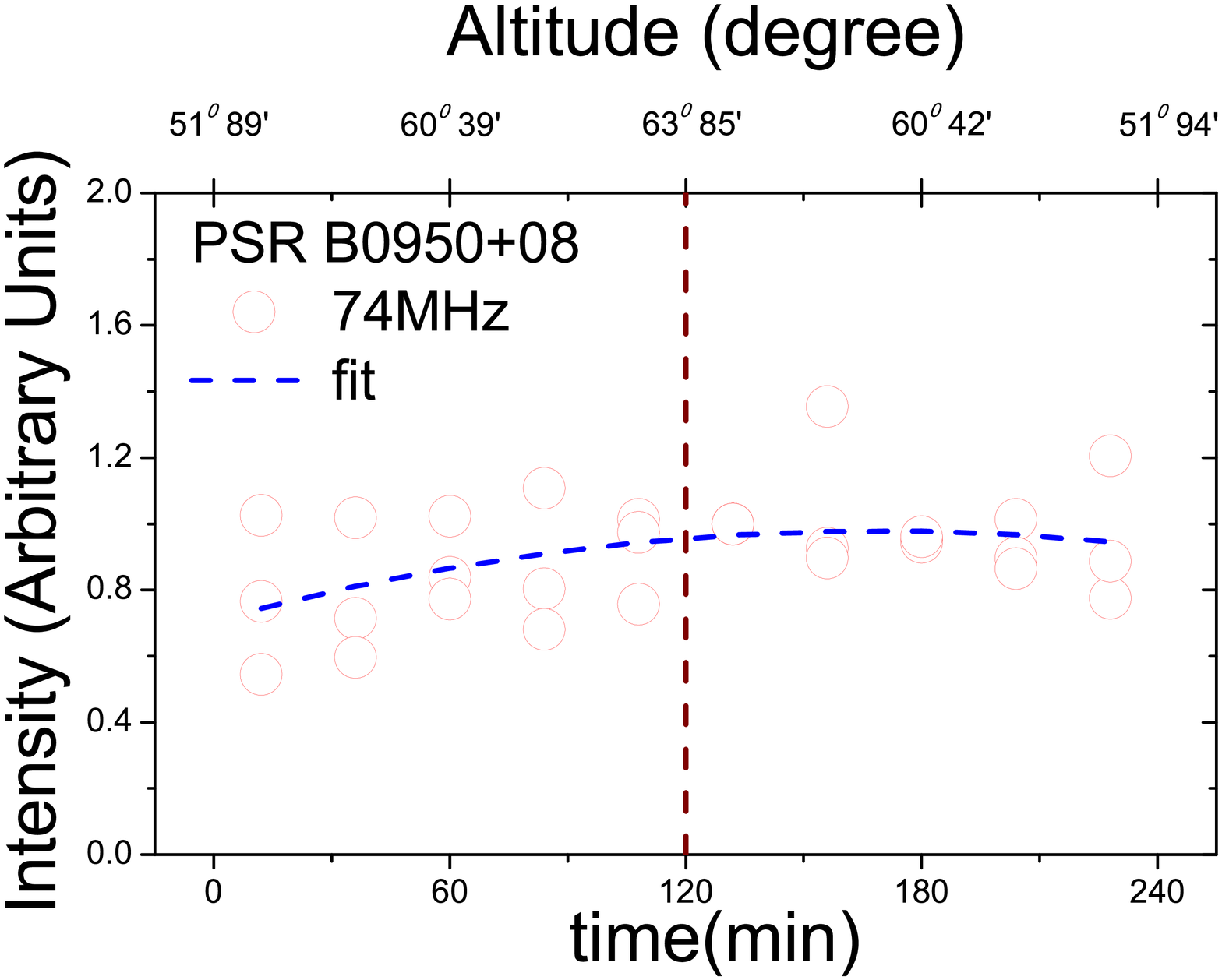}\\
\caption{The flux density at different altitudes at two frequencies.  PSR~B0950+08 passed the meridian at 120 minutes after the first observing frame, with an altitude = 63.85$^\circ$.  The two panels demonstrate the sensitivity dependence on a target's zenith angle and the observing frequencies of LWA1 in beam-tracking mode.  We corrected for these systematic effects when calculating the intensity of GPs. Each data point is the 24-minute averaged intensity.  There are three observations from different dates in each panel.
}
\label{altitudecorr}
\end{center}
\end{figure}

The sensitivity of LWA1 is also dependent on the altitude of the target.  To correct for this effect we calculated sequences of 24~minute averages of the pulse flux densities and fit the variation with a polynomial function, with a maximum value of unity at or near the meridian, as shown in Figure \ref{altitudecorr}.  Then the flux of any particular pulse was divided by the polynomial to remove the zenith angle dependence for future analysis.

\subsection{Profile of PSR~B0950+08}\label{profile0950}
An adequate fit to the pulse profiles required three Gaussian components, as demonstrated in Figure \ref{phasesingledouble0950}.  While comparing the AP profile at two frequencies, we noted that the leading peak position does not shift more than a temporal bin, while the second peak's shift is more obvious.

\begin{figure}
\begin{center}
\includegraphics[width=0.75\textwidth]{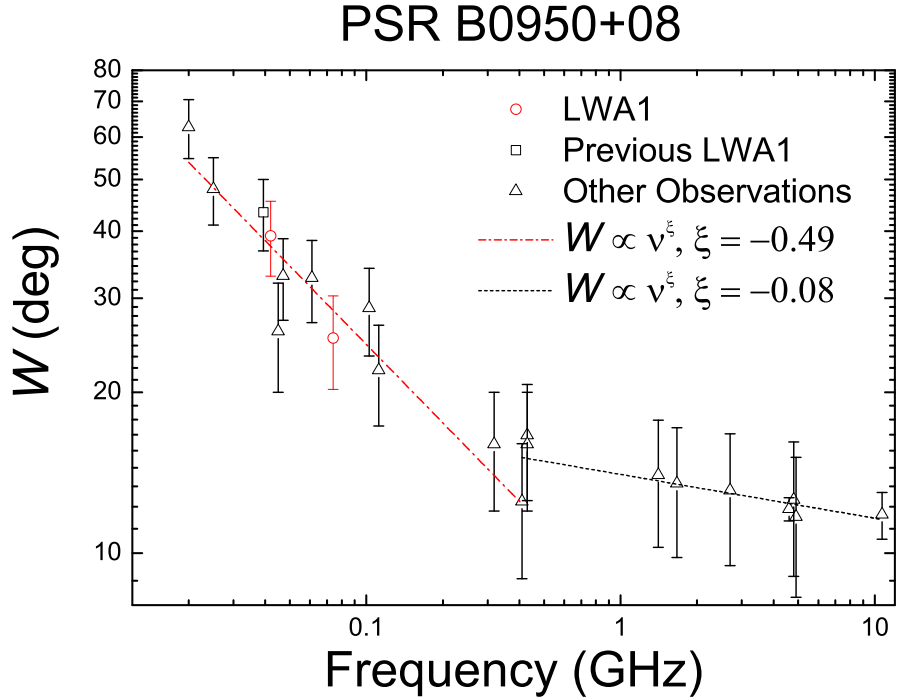}
\caption{The spectra of pulse widths observed for PSR~B0950+08 normalized to the period of $360^{\circ}$.  The pulsar's period is 253.06~ms.  We fit the spectral index for two frequency ranges, separated at 410~MHz.  The vertical error bar is one sigma assuming Poisson statistics unless otherwise specified in references for the observations cited.  Observations are from 
\citet{2015AJ....149...65T}:39.4~MHz; 
\citet{1971ApJS...23..283M}:0.41~GHz, 1.665~GHz; 
\citep{1975A&A....38..169S}:2.7~GHz, 4.9~GHz; 
\citet{1979SvA....23..179I}: 61~MHz, 102.5~MHz; 
\citet{1981AJ.....86..418R}: 430~MHz; 
\citet{1986A&A...161..183K}: 4.6~GHz, 10.7~GHz; 
\citet{1992ApJ...385..273P}: 25~MHz, 47~MHz, 112~MHz, 430~MHz, 1408~MHz, 4800~MHz; 
\citet{1995A&A...301..182R}: 45~MHz 
and this work: 42~MHz, 74~MHz.
}  
\label{pulsewidthindex}
\end{center}
\end{figure}

The width of the AP measured at the half maximum height from the leading to trailing components, $W$, is 25.6 ms and 17.8 ms at 42 and 74~MHz respectively. Figure \ref{pulsewidthindex} shows $W$ for frequencies above 20~MHz.  We find a temporal broadening spectral index of $\xi = -0. 49\pm0.064$ when our observations are fit along with other observations of this pulsar at frequencies from 20 to 410 MHz, as shown in Figure \ref{pulsewidthindex}.  The spectral index seems to change at about 0.5~GHz.

\begin{figure}
\begin{center}
\includegraphics[width=0.75\textwidth]{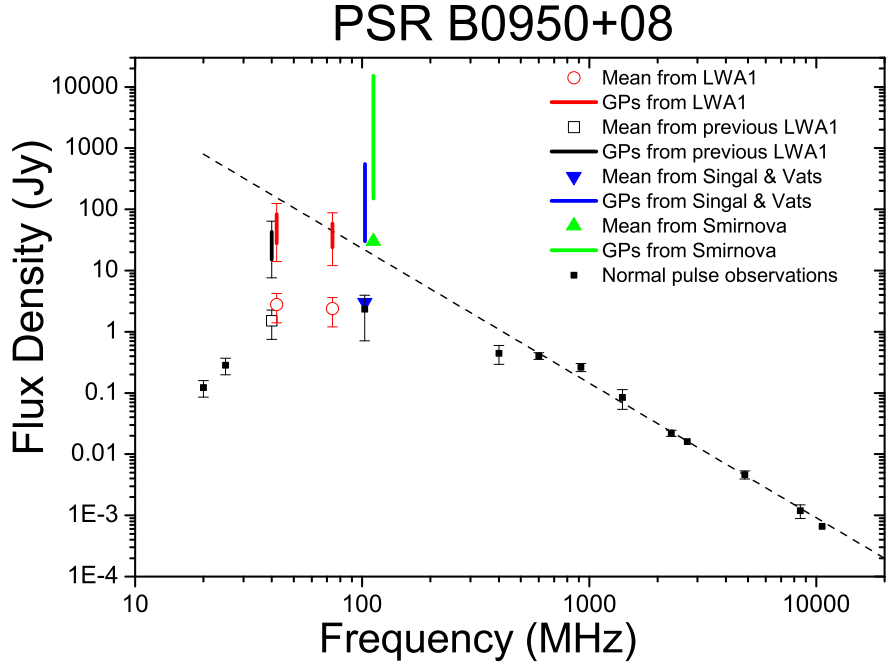}
\caption{Observed flux densities for PSR~B0950+08 for different observing frequencies are shown here.  The AP flux density data points have 50\% error bars unless otherwise specified in references for the observations cited.  The vertical line indicates the range of GP values.  The thin error bars extending above and below indicate the 50\% error associated with the lowest GP and highest GP flux density.  The black solid square represents the flux density for the AP.  Red halo circles are the AP flux densities for LWA1 at 42 and 74~MHz (this work).  The thick red vertical line indicates the range of GPs' values.  The black halo square represent previous LWA1 observation at APs and GPs by \citet{2015AJ....149...65T} at 39.4~MHz.  Blue down triangle and green up triangle similarly represent observations by \citep{2012AJ....144..155S} at 103~MHz, and by \citet{2012ARep...56..430S} at 112~MHz respectively.  The AP observations (black squares) are from (left to right: points 1--3, \citep{2013MNRAS.431.3624Z}; points 4--6, and 9, \citep{GouldLyne}; point 7, 10, and 12, \citep{Seiradakis}; points 8 and 11, \citep{vonHoensbroechXilouris}.  The dashed black line is a fit through the AP observations, of spectral index $-$2.2.
} 
\label{fluxspectra0950}
\end{center}
\end{figure}

The period-averaged flux densities are 2.8 Jy and 2.4 Jy at 42 and 74~MHz, respectively.  Figure \ref{fluxspectra0950} shows flux densities for APs at frequencies of 20~MHz and above, along with observations of GPs by \citet{2012AJ....144..155S} at 103~MHz and \citet{2012ARep...56..430S} at 112~MHz.  In addition, we have added data points from this work.  The error bar on our AP flux densities indicates our $\sim$50\% uncertainty.  The figure shows that both APs and GPs have lower flux density below $\sim$100~MHz.

\subsection{GPs from PSR~B0950+08}\label{gps0950}
\begin{figure}
\begin{center}
\includegraphics[width=0.75\textwidth]{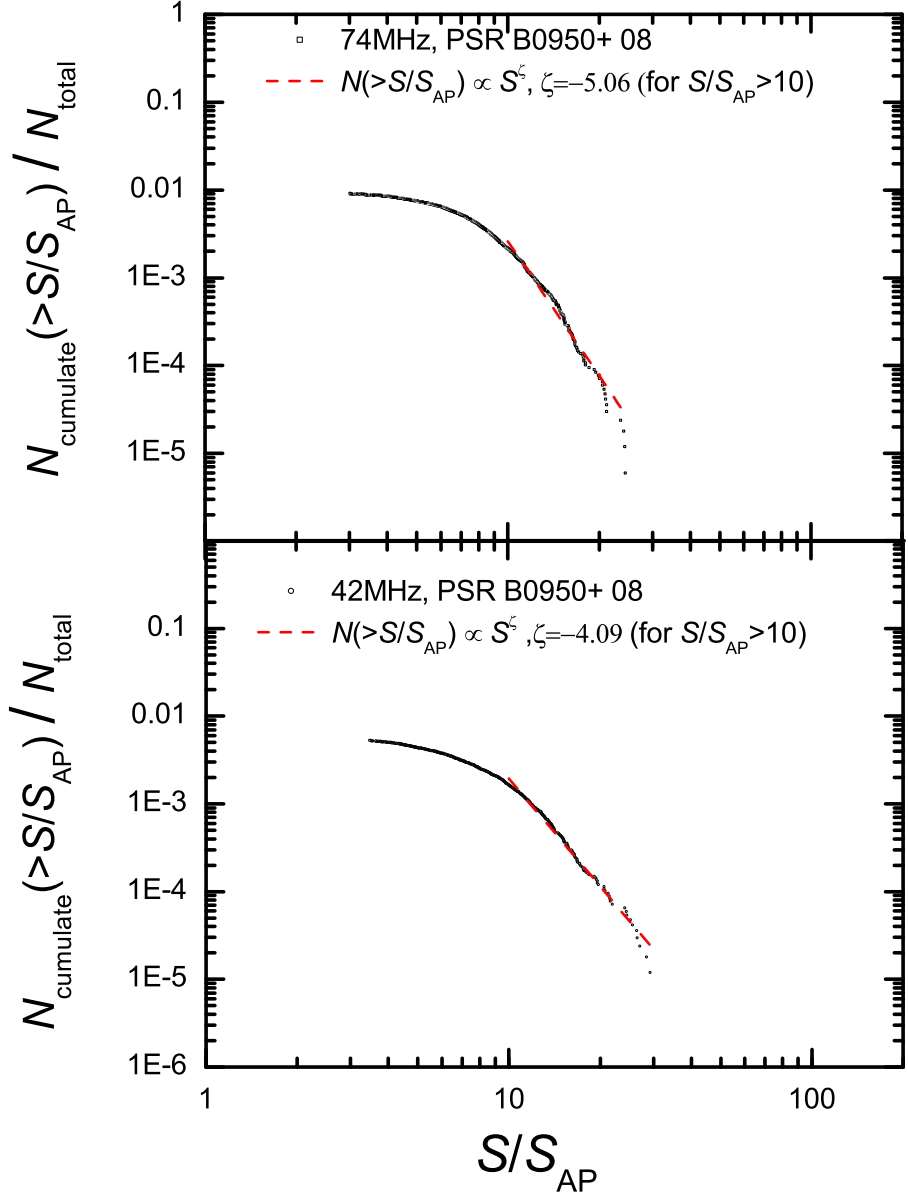}
\caption{The cumulative pulse number $N(>S/S_{\rm AP}$) plotted versus flux density.  The red dashed line in both panels is the power-law fit for $S/S_{\rm AP}>$10.
}
\label{loglogGpsEventRate0950}
\end{center}
\end{figure}

The cumulative distribution of pulses with flux density 10 times larger than that of the AP ($S_{\rm AP}$) was found to follow a power law $N(>S/S_{\rm AP})\propto S^\zeta$, where $\zeta=-4.09$ and $-5.06$ for 42 and 74~MHz respectively, as shown in Figure \ref{loglogGpsEventRate0950}.  This supports our use of the definition of GPs as 10 times the AP's flux density for PSR~B0950+08 as found in \citet{2012AJ....144..155S}.  The strongest GP at 42~MHz had a flux density 29 times larger than the APs, while at 74~MHz the strongest one was found to be 24 times larger.

The cumulative distribution of pulse strength for 42 and 74~MHz has a steeper power-law for the GPs than found by \citet{2012AJ....144..155S} for 103~MHz and \citet{2012ARep...56..430S} for 112~MHz.  All these results indicate that PSR~B0950+08 is weaker, and produces less frequent and less intense GPs at low frequency than at 100~MHz.  All GPs have double peaked structures and the phase of GPs drifts through the range of the AP profile, as shown in Figure \ref{phasesingledouble0950}.

\begin{figure}
\begin{center}
\includegraphics[width=.75\textwidth]{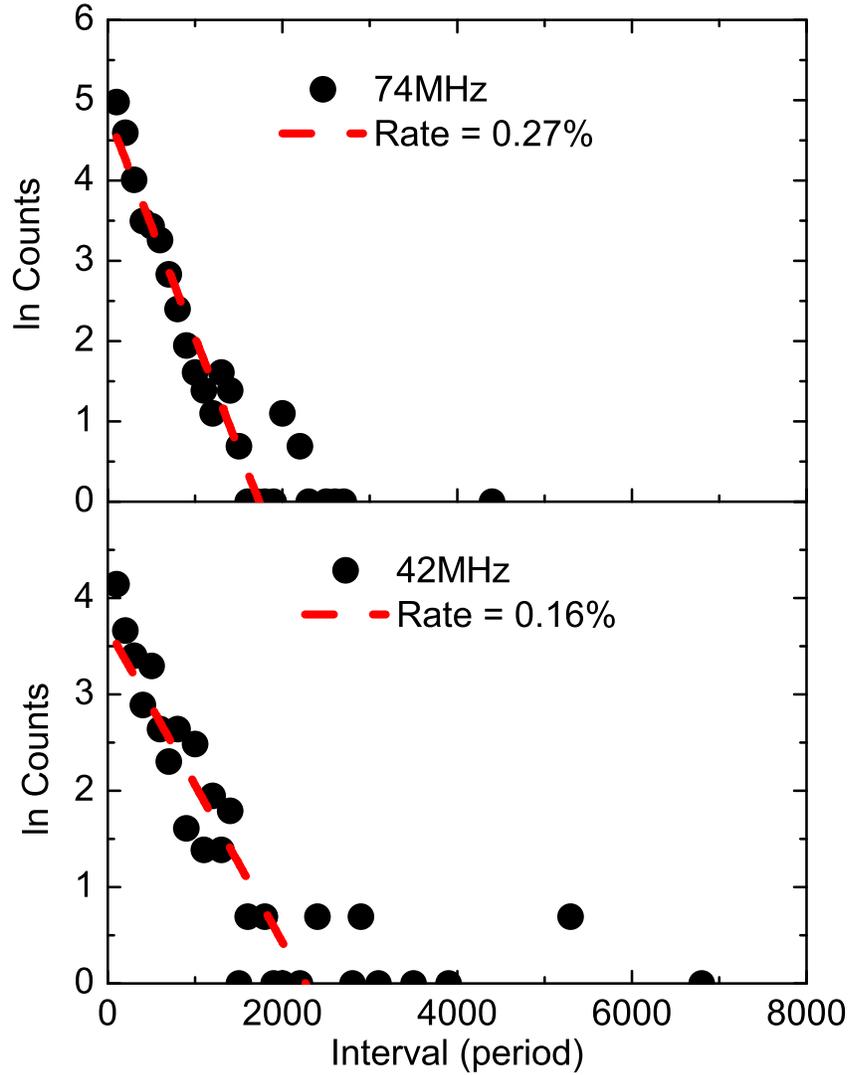}
\caption{GP interval distributions at both frequencies. Intervals on the horizontal axis are in units of 100 periods. The rate shown is the slope of a fit from 0 to 2800 periods and 0 to 1900 periods for 42 and 74~MHz, respectively. We did not find interesting clustering or periodic behaviors regardless of the chosen binning interval.
}
\label{intervaldistribution}
\end{center}
\end{figure}

If each GP is an independent event with a uniform probability for pulse generation at any given time, we expect the distribution of intervals between GPs to exhibit an exponential functional dependence. For simplicity, we will use the number of accumulated pulse periods, $p$, as a time measure. Then, the interval distribution for a single event as a function of the number of periods is given by
\begin{equation}
I_1(p) = r\ \mbox{exp}(-r p),
\label{eqn:exponential distribution}
\end{equation}
where $r$ is the average event rate (GPs per period). 
The observed distribution is shown in Figure \ref{intervaldistribution}, which indicates that the GPs are indeed independent events.

\section{Scattering Analysis}\label{sect:scattering}
We made use of the CLEAN-based algorithm \citep{2003ApJ...584..782B} to analyze the scattering effect on the temporal broadening of the GP and AP from the PSR~B0950+08.  The observed pulses are composed of the intrinsic pulse convolved with propagation effects and instrumental response.  The CLEAN-based algorithm utilizes an accumulated delta-like signal to restore the intrinsic pulse.  This approach allows for the deconvolution of various profile shapes without knowledge of the intrinsic profile.  The recorded signal $P_{obs}$ is
\begin{equation} 
P_{obs}(t) = I(t)\otimes pbf(t)\otimes r(t),
\label{convolutions}
\end{equation}
where $I(t)$ is the intrinsic pulse profile, $pbf(t)$ is the pulse-broadening function and $r(t)$ is a function which gives the combined instrumental responses including effects due to data reduction.

\begin{figure}
\begin{center}
\includegraphics[height=.75\textwidth]{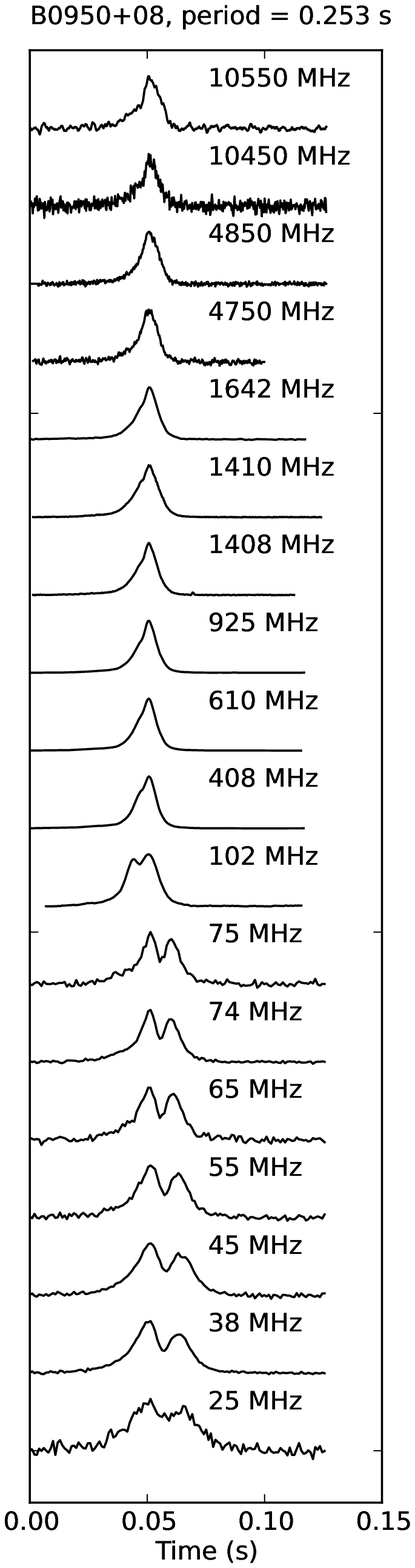}
\caption{Profiles for different observing frequencies.  The data with frequency range from 25 to 75~MHz are from LWA Pulsar Data Archive \citep{2014arXiv1410.7422S}, the rest are from European Pulsar Network (EPN) data archive with contributors:\citet{1998MNRAS.301..235G, 1994ApJ...422..671A, 1997A&AS..126..121V, 1997ApJ...488..364K, 1997A&A...322..846K}, and \citet{1995A&AS..111..205S}.  Profiles are aligned with the phase of the maximum. 
}
\label{figallB0950+08}
\end{center}
\end{figure}

We applied this algorithm to data from the LWA Pulsar Data Archive \citep{2014arXiv1410.7422S} for observations from 25 to 75~MHz and the EPN data archive \citep{1998MNRAS.301..235G,1994ApJ...422..671A,1997A&AS..126..121V,1997ApJ...488..364K,1997A&A...322..846K,1995A&AS..111..205S} for 102~MHz to 10.6~GHz, and this work.  We include as part of the instrumental responses the effect derived from setting the bin size of the time series to be same as the temporal resolution and a similar effect for the frequency channel width.  For simplicity, we compare only the scattering time constant from a thin screen model for APs in Figure \ref{figallB0950+08} with results for pulses with SNR$>7$.

\begin{figure}
\begin{center}
\includegraphics[height=.75\textwidth]{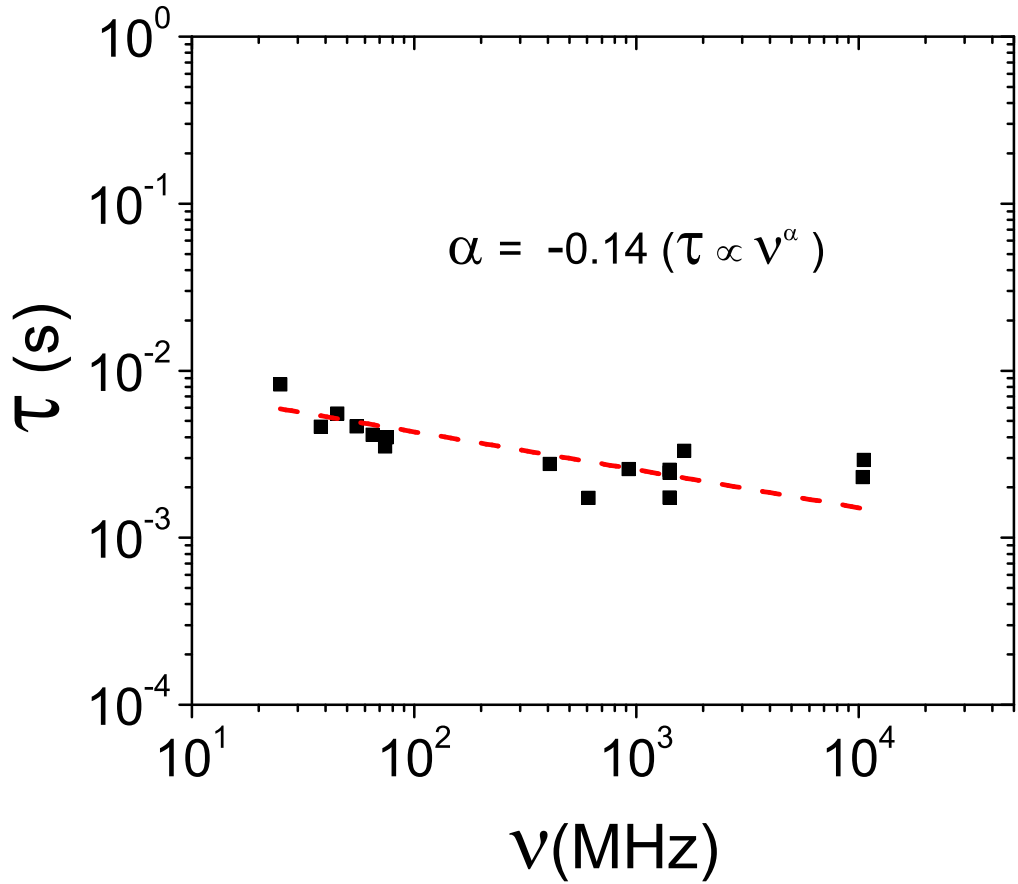}
\caption{The exponential decay time $\tau$ of the profiles derived from the CLEAN-based method, assuming a thin screen model.  The derived spectral index is $\alpha = -0.14\pm$0.03 for $\tau\propto\nu^\alpha$. Data used here are as same as in Figure \ref{figallB0950+08}.
}
\label{B0950+08spectralindex}
\end{center}
\end{figure}

After deconvolution we found the  exponential decay time in the thin screen scattering module from pulses with SNR$>7$ centered at 42 and 74~MHz indicate a frequency  scaling  index of $-1.45\pm0.14$ which differs to $-0.91\pm0.2$ calculated from the APs. And over a much wider range of frequencies, see Figure \ref{figallB0950+08} and \ref{B0950+08spectralindex}, is $-0.14\pm0.03$, which also greatly differs from of pulses with SNR$>7$.

However, scatter-broadening should be a property of the ISM through which the pulse is traveling, and therefore should be the same for all types of pulses from a given pulsar.  So the inconsistent exponential decay time we had from pulses with SNR$>7$ (2.5$\pm1.3$/1.1$\pm0.7$~ms for 42/74~MHz) and AP (5.4/3.4~ms for 42/74~MHz) should be resulted from the dominated evolution of the profile at different frequencies, such as radius-to-frequency mapping \citep{1978ApJ...222.1006C}, rather than the scattering effect from ISM.  Also as \citet{2013MNRAS.434...69L,2015ApJ...804...23K} argue one can obtain reliable values of scatter time (from any method) only when the scatter time is significantly larger than the width of the profile. The exponential decay time derived from CLEAN method is much smaller than the profile width.

So the ISM scattering effect still is not strong enough to be distinguished when the frequency is down to 42~MHz within line of sight to PSR~B0950+08.

\section{Pulse Emission Altitude}\label{sect:altitudes}

Observations of GPs can be used to probe a pulsar's emission region. This can be done in a more subtle fashion if simultaneous observations at different frequencies are taken for a given pulse.  Of particular importance is the observed separation in time between the two peaks in a GP or AP profile at the two observing frequencies. We consider the difference in the the altitude of emission at the two observing frequencies.  We do this both for APs and GPs and find a similar difference in emission altitude between the two frequencies.  In analyzing the difference in the altitude of the emission region for the two frequencies observed we made use of the dipolar magnetic field model, as discussed in \citet{2012hpa..book.....L}. We will briefly summarize the basic features of this model that were used to calculate the altitude difference.

\begin{figure}
\begin{center}
\includegraphics[width=.75\textwidth]{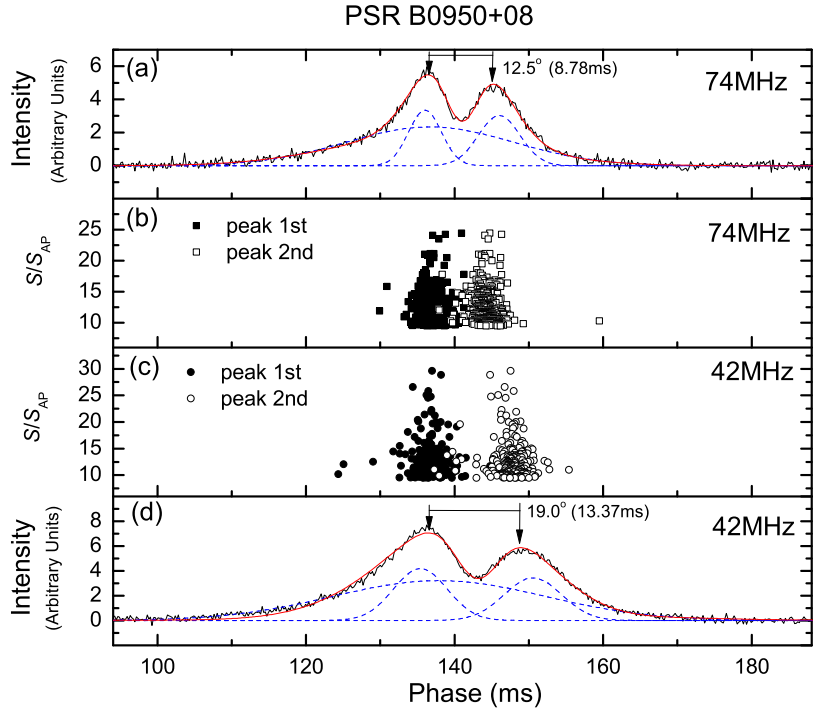}
\caption{The AP profiles and GP peak phases at two frequencies.  Panels (a) and (d) show the profile of the AP at each frequency, the three dash lines are Gaussian fit to the profile using a least squares fitting algorithm. The red solid curve is the summation of three Gaussians.  Panels (b) and (c) show GP peak phases and flux density at each frequency compared to the flux density of the AP.  A GP is defined as a pulse with flux density ratio ($S/S_{\rm AP}$)$>$10 in this study.  The pulsar's period is 253.06~ms.
}
\label{phasesingledouble0950}
\end{center}
\end{figure}

In the dipolar magnetic field model the field geometry is described using polar coordinates such that the ratio between $\sin^2 \theta$ and a radius ${r}$ is fixed along a given magnetic field line.  So, we can connect a point at $(r, \theta)$ along the last open field line of a dipole at the radius of light cylinder $r_{LC}$ by
\begin{equation} 
\frac{\sin^2\theta}{r} = \frac{\sin^2 90^\circ}{r_{LC}}.
\label{dipoleradius1}
\end{equation}
The opening angle $\rho$ is the half-width of the beam; it can be related to $\theta$ as 
\begin{equation} 
\tan \theta = -\frac{3}{2\tan\rho}\pm \sqrt{2+\frac{9}{4 \tan^2\rho}}.
\label{dipoleradius2}
\end{equation}
if the emission cone is confined by the last open magnetic field line \citep{2001ApJ...555...31G}. From equations (\ref{dipoleradius1}) and (\ref{dipoleradius2}), the opening angle $\rho$ can be related to the emission height $r$ as
\begin{equation} 
\rho = 86^\circ \left(\frac{r}{r_{LC}}\right)^\frac{1}{2},
\label{dipoleradius3}
\end{equation}
using the small-angle approximation for both $\rho$ and $\theta$. The opening angle is also related to the observed pulse component separations $\Delta\phi$ as \citep{1984A&A...132..312G}
\begin{equation} 
\cos \rho = \cos \eta \cos (\eta+\psi) + \sin(\eta)\sin(\eta+\psi)\cos\left(\frac{\Delta\phi}{2}\right),
\label{dipoleradius4}
\end{equation}
where $\eta$ is the angle between rotation and magnetic axes, and $\psi$ is the impact parameter, which is the angle of closest approach of the magnetic axis and the line-of-sight to the observer.  Given the component separations $\Delta\phi=19^\circ\pm0.75^\circ$ and $12.5^\circ\pm0.54^\circ$ at 42 and 74~MHz respectively for the APs from Figure \ref{phasesingledouble0950} and $\eta=5.9^\circ$ and $\psi=-4.2^\circ$ for PSR~B0950+08 from \citet{1988MNRAS.234..477L}, we find that the emission altitudes of both frequencies are $r$(42~MHz) = 29267$\pm36$~m ($\approx0.242\%$ of $R_{LC}$) and $r$(74~MHz) = 29013$\pm17$~m ($\approx0.240\%$ of $R_{LC}$), and the altitude difference is 254$\pm53$ m.  For GPs, given the component separations as 15.2$^\circ\pm0.30^\circ$ and 10.3$^\circ\pm0.13^\circ$ for 42 and 74~MHz respectively, the altitude is 29106$\pm11$~m ($\approx0.241\%$ of $R_{LC}$) and 28950$\pm3$~m ($\approx0.240\%$ of $R_{LC}$) for 42 and 74~MHz respectively, and the altitude difference is 156$\pm14$~m.  Because the uncertainties are not available on $\eta$ and $\psi$ from \citet{1988MNRAS.234..477L}, the uncertainties noted above in altitudes $r$ are solely calculated from uncertainties of $\Delta \phi$.  The emission altitudes of APs are within generally accepted $10\%$ of the $R_{LC}$ \citep{2003A&A...397..969K} and are much closer to the surface of the pulsar than to the light cylinder. The mean emission altitudes of GPs are a little closer to the surface of the pulsar than APs for both 42 and 74~MHz.

There are studies shows a variety ratios of emission altitude to light cylinder of GPs, such as GPs of PSR~J1823$-$3021A are emitted no higher than 4 km above ordinary emission \citep{2007MNRAS.378..723K}, or GPs of PSR~B1821$-$24A are speculated to have emission altitudes near the light cylinder where is much different from the regions of radio emission is placed traditionally\citep{2015ApJ...803...83B}.

As we can see from panels (b) and (c) in Figure \ref{phasesingledouble0950}, the weaker GPs have broader range of phases.  The stronger GPs appear to concentrate toward the maximum intensities of components.  This was also indicated by \citet{2006ARep...50..915S} with analysis of this pulsar at 111~MHz. Note that the mean peak separation is less for the GPs than the APs.

\section{Conclusion}\label{sect:conclusion}

We observed PSR~B0950+08 simultaneously at 42 and 74 MHz detecting GPs and APs. Our study shows GPs and APs from PSR~B0950+08 are much weaker than at $\sim$100~MHz. This indicates a turn-over in the spectrum of GPs between 74-100 MHz which is consistent with previous observations \citep{2015AJ....149...65T}. We analyzed the effect of scattering using the CLEAN-based algorithm \citep{2003ApJ...584..782B} and found that the scattering effect still can not be observed for frequency down to 42~MHz. There are other effects that dominated the profile broadening, rathe than the scattering effect from the ISM.  We determined the emission altitude in the context of a dipolar magnetic field for our set of observations for both GPs and APs.  We found $r$(42~MHz) = 29.27~km ($0.242\%$ of $R_{LC}$) and $r$(74~MHz) = 29.01~km ($0.240\%$ of $R_{LC}$) for the average pulse, while for giant pulses, $r$(42~MHz) = 29.10~km ($0.241\%$ of $R_{LC}$) and $r$(74~MHz) = 28.95~km ($0.240\%$  of $R_{LC}$).
The difference in emission altitudes of GPs is similar to that of the APs.  Most GPs can only be detected at one frequency which implies that the emission of GPs is localized within a small region and drifts within the emission region of the APs.  The average component separation of the GP is smaller than that of the AP, and the GP's component separations are closer to the average with larger intensity.

\section*{Acknowledgments}
We acknowledge insightful discussions with Kevin Stovall and Roger Link.  Construction of the LWA has been supported by the Office of Naval Research under Contract N00014-07-C-0147.  Support for operations and continuing development of the LWA1 is provided by the National Science Foundation under grant AST-1139974 and AST-1139963 of the University Radio Observatory program.  The computation is supported by the Advance Research Center of Virginia Tech.  Data reduction was performed using the BlueRidge system at Virginia Tech.

{\it Facility:} \facility{LWA}


\end{document}